\documentclass[conference, a4paper,UTF8]{IEEEtran}
\IEEEoverridecommandlockouts
\usepackage{fancyhdr} 
\usepackage{color}
\usepackage{amssymb}
\usepackage{amsmath}
\usepackage{amsmath,amssymb,amsfonts}
\usepackage{algorithmic}
\usepackage{graphicx}
\usepackage{textcomp}
\usepackage{amsmath,amssymb,amsfonts}
\usepackage{algorithmic}
\usepackage{graphicx}
\usepackage{float}
\usepackage{subfigure}
\usepackage{setspace}
\usepackage{hyperref}
\usepackage{textcomp}
\usepackage{amsfonts}
\usepackage{paralist}
\usepackage{tikz}
\usetikzlibrary{shapes.arrows}
\usepackage{amsmath}
\usepackage{algorithm}
\usepackage{algorithmic}
\usepackage{amsthm}
\usepackage{warpcol}
\usepackage{booktabs}
\usepackage{multirow}
\usepackage{graphicx}
\usepackage{booktabs}
\usepackage{rotating}
\usepackage{placeins}
\usepackage{mathrsfs}
\usepackage{ifsym}
\usepackage{tikz} 
\usepackage{subfigure}
\usepackage{listings}

\usepackage{geometry}
\usetikzlibrary{decorations.pathreplacing}
\usetikzlibrary{decorations.pathreplacing,calligraphy}
\renewcommand\figurename{Figure}      
\renewcommand\contentsname{Contents}   
\renewcommand\abstractname{Abstract}   
\newcommand{\tabincell}[2]{\begin{tabular}{@{}#1@{}}#2\end{tabular}}%
\usepackage{makecell}
\renewcommand\tablename{Table}            
\usepackage{lastpage}      
\usepackage{setspace}          
\usepackage{algorithm} %format of the algorithm 
\usepackage{algorithmic} %format of the algorithm 
\usepackage{multirow} %multirow for format of table 

\usepackage{amsmath,bm} 
\usepackage{xcolor}                         
\usepackage{layout}  
\usepackage{amsmath}   
\usepackage{hyperref}
\usepackage{color}
\usepackage{amsfonts}
\usepackage{graphicx}       
\usepackage{hyperref}                                  
\begin{document}
\begin{spacing}{1}

\title{Evaluation and Comparison of Diffusion Models with Motif Features}
\author{
\IEEEauthorblockN{Fangqi Li}
\IEEEauthorblockA{\textit{School of Cyber Science and Engineering, SEIEE, SJTU}\\
\IEEEauthorblockN{$\{$solour\_lfq$\}$@sjtu.edu.cn}}
}
\date{Today}
\maketitle
\begin{abstract}
Diffusion models simulate the propagation of influence in networks. 
The design and evaluation of diffusion models has been subjective and empirical. 
When being applied to a network represented by a graph, the diffusion model generates a sequence of edges on which the influence flows, such sequence forms a temporal network. 
In most scenarios, the statistical properties or the characteristics of a network are inferred by analyzing the temporal networks generated by diffusion models. 
To analyze real temporal networks, the motif has been proposed as a reliable feature. 
However, it is unclear how the network topology and the diffusion model affect the motif feature of a generated temporal network. 
In this paper, we adopt the motif feature to evaluate the temporal graph generated by a diffusion model, thence the diffusion model itself. 
Two benchmarks for quantitively evaluating diffusion models with motif, stability and separability, are proposed and measured on numerous diffusion models. 
One motif-based metric is proposed to measure the similarity between diffusion models. 
The experiments suggest that the motif of a generated temporal network is dominated by the diffusion model, while the network topology is almost ignored. 
This result indicates that more practical and reliable diffusion models have to be designed with delicacy in order to capture the propagation patterns of real temporal networks. 

\textbf{Keywords:} network analysis, diffusion model, motif.

\end{abstract}
\section{Introduction}
\label{section:1}
To capture and analyze the dynamics of the influence propagation in a network, it is necessary to record the time at which the influence passes through an edge. 
Therefore the temporal network, whose edge has an additional time stamp attribute, has become an research target of interest.
A temporal network can be formulated by $G=\left\{V,E\right\},$
where $V$ is the collection of all participants/nodes of the network, $E$ is the collection of edges.
To record time-dependent properties in temporal networks, $E$ is a subset of $\left(V\times V\times T \right)$, where $T$ is the domain of time. 
An edge denoted by $e=(u,v,t)\in V$ implies that the influence is propagated from $u$ to $v$ at time $t$, i.e., $e$ is activated at time $t$. 

By recording the process of spreading the influence/information in a community, the temporal network provides a comprehensive description of time-dependent network structures such as social networks, traffic networks, communication networks, logistics networks, etc.
Such comprehensiveness paves the way to many studies on the time-dependency of networks \cite{9169627}\cite{8714706}\cite{7925090}. 

As a complex object in itself, temporal networks can hardly be compared and analyzed without abstract features.
Among them a flagship option is the \emph{motif feature}. 
A motif is a tiny propagation pattern within a temporal network. 
The motifs with three nodes and three edges is of particular interest since they can be efficiently counted and they reflect the network's characteristics. 
To form the motif feature map of a temporal network, all three/two-node, three-edge, $\delta$-temporal (all three edges have to be activated within a time window of width $\delta$) motifs are counted, forming a feature vector with 36 components. 
It has been demonstrated that such feature is representative for temporal networks \cite{paranjape2017motifs}\cite{benson2016higher}\cite{abuoda2019link}. 

However, temporal network data is scanty compared to static networks, which can be compactly saved as an ordinary graph. 
It is expensive to maintain a temporal network database, not to mention the concerns on privacy and security.
Moreover, many mainstream studies model the influence propagation as a stochastic process. 
But a real temporal network is only an sampled instance from the underlying probability space. 
Therefore, it is always necessary to generate temporal networks from the static network topology.
This is usually done by applying a diffusion model to a static network. 
A diffusion model defines how influence flows among nodes within a graph, i.e., which edges are to be activated at a certain time stamp. 

Designing diffusion models is a challenge as well as an artwork. 
In order to accomodate to specific tasks, the diffusion models have to meet some mathematical properties such as \emph{submodularity} for influence maximization. 
Recent options in simulating information propagation and predicting links such as \emph{graph neural networks} often sacrifice the temporal information for efficiency in deducing the influence range \cite{wang2020model}\cite{yu2019rum}. 
Therefore they fail to befit studies that focus on temporal properties. 

It is generally hard to quantitively compare diffusion models, not to mention evaluating them. 
However, we conjecture that the motif features might shed light on this problem. 
Intuitively, the artificial temporal network generated from a diffusion model should appear as a real temporal network from the motif' s perspective. 
If the temporal network generated by initializing the diffusion model with different random seeds appear to be distinguishable according to their motif features then the diffusion model is not \emph{stable}.
If the temporal network instances generated by running the same diffusion model on different network structures appear to be identical w.r.t. motif then the diffusion model cannot \emph{separate} different underlying topologies.

Meanwhile, the deterministic factor of the motif feature of a generated temporal network remains unclear.
So it is crucial to measure how the network topology and the diffusion model cooperate to form the motif features. 
We formally address these two properties and conduct comprehensive experiments to evaluate the influence of the diffusion model and the network topology to the motif features. 

The contributions of this paper are three-fold:
\begin{enumerate}
\item We propose two motif-based quantitive metrics to measure the performance of diffusion models, namely \emph{stability} and \emph{separability}.
They are boiled down to tractable statistics, making the evaluation of diffusion models more practical.
\item A motif-based distance measure is proposed to calculate the similarity between diffusion models. 
\item We evaluate and compare several classical diffusion models using the proposed metrics. 
The results challenge previous studies that built their diversity solely in network datasets and call for designing more distinctive diffusion models. 
\end{enumerate}

The preliminaries and formulations are collected in Section \ref{section:2}. 
The motivations and the designed metrics are presented in Section \ref{section:3}. 
Experimental results are collected in Section \ref{section:4}, Section \ref{section:5} concludes the paper.  

\section{Preliminaries and Formulations}
\label{section:2}
\subsection{Diffusion Model}
The topology of a network is a graph, optionally with direction and weight:
\begin{equation}
E\subset\left(V,V \right)\text{ or }\left(V,V,\mathbb{R} \right).
\end{equation}
A diffusion model $M$ is a probabilistic mapping from a subset $S$ of the collection of nodes, known as the \emph{seeds}, to another destination subset $D$ of $V$.
\begin{equation}
M:D\leftarrow S,
\end{equation}
where $S\subset D\subset V$. 
$M$ usually returns $D$ as a cascade that consists of an ordered sequence of activated edges. 

For example, the independent cascade model, $M_{\text{IC}}$, operates as follows:
\begin{enumerate}
\item Activate a subset $E'\subset E$, with
\begin{equation}
\text{Pr}\left\{e\in E' \right\}=\phi(e),
\end{equation}
where $\phi(\cdot)$ is a monotonical function that maps the weight of $e$ to $[0,1]$. 
\item For a node $v\in V$, if there is a path from $v$ to $S$ in $E'$ then $v$ is added to $D$. 
\end{enumerate}

A threshold model $M_{\text{threshold}}$ drops the independence between edges:
\begin{enumerate}
\item Define a topological order on $V$:
\begin{equation}
T:v_{1},v_{2},\cdots,v_{|V|}.
\end{equation}
\item Set $D=S$.
\item Iterate over $T$, if the parents of $v\in V$ satisfies $v$' s activation threshold condition then $v$ is added to $D$.
\end{enumerate}
For example, the threshold condition for linear threshold model takes the form:
\begin{equation}
\sum_{u,e=(u,v,w_{e})\in\mathcal{E}}\mathbb{I}[u\in D]\cdot w_{e}\geq h_{v},
\end{equation}
where $w_{e}$ is the weight of edge $e$ and $h_{v}$ is a node-dependent threshold. 

More delicate diffusion models take the interaction among the parents of a node to compute its probability of being activated \cite{bao2016inferring}.
These choices are usually computationally expensive, especially for large networks.

Diffusion model is crucial in addressing optimization tasks regarding influence propagation, such as marketing, advertising, etc. 
As for real temporal networks, heterogeneous messages are passed across the network simultaneously.
Hence to forge a real temporal network, we combine several independent cascades into one generated temporal network instance.
The diffusion model $M$ is run on a static network $G$ for $P$ times, resulting in $P$ independent sequences of edge activation:
\begin{equation}
\mathcal{P}=\left\{(e_{p,1},e_{p,2},\cdots,e_{p,l_{p}}) \right\}_{p=1}^{P}.
\end{equation} 
These $P$ sequences are then combined to form a generated temporal network $TG_{M,G}$ by Algo. \ref{algorithm:1}. 

The underlying static network $G$ and $TG_{M,G}$ share the same collection of nodes $V$. 
\begin{algorithm}[htbp] 
\caption{Edge combination: $TG_{M,G}\leftarrow \mathcal{P}. $}  
\label{algorithm:1}  
\begin{algorithmic}[1]
\REQUIRE $\mathcal{P}$.
\ENSURE $TG_{M,G}$.
\STATE $\text{time}=0$;
\WHILE {$\mathcal{P}$ not empty}
\STATE Sample a sequence $E'$ from $\mathcal{P}$;  
\STATE Pop the first term $e$ from $E'$;
\IF {$E'$ is empty.}
\STATE Delete $E'$ from $\mathcal{P}$.
\ENDIF
\STATE $t\sim\text{Exponential(3)}$;
\STATE $\text{time}+=t$;
\STATE Add $(e,\text{time})$ as an edge of $TG_{M,G}$;
\ENDWHILE
\end{algorithmic}  
\end{algorithm} 
Where a random variable subject to an exponential distribution is adopted to simulate the intervals between events as in queuing theory and communications \cite{giambene2005queuing}. 

\subsection{Motif}
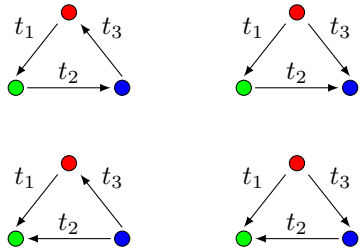
\begin{figure}[htbp]
\centering
\begin{tikzpicture} 
\draw (0,0) circle (0.1) [fill=red];
\draw (-0.7,-1) circle (0.1) [fill=green];
\draw (0.7,-1) circle (0.1) [fill=blue];
\draw (-0.15,-0.15)--(-0.7,-0.85) [-latex];
\draw (-0.55,-1)--(0.55,-1) [-latex];
\draw (0.7,-0.85)--(0.15,-0.15) [-latex];
\node at (-1.15/2,-0.2) {$t_{1}$};
\node at (0,-0.8) {$t_{2}$};
\node at (1.15/2,-0.2) {$t_{3}$};

\draw (0+3,0) circle (0.1) [fill=red];
\draw (-0.7+3,-1) circle (0.1) [fill=green];
\draw (0.7+3,-1) circle (0.1) [fill=blue];
\draw (-0.15+3,-0.15)--(-0.7+3,-0.85) [-latex];
\draw (-0.55+3,-1)--(0.55+3,-1) [-latex];
\draw (0.15+3,-0.15)--(0.7+3,-0.85) [-latex];
\node at (-1.15/2+3,-0.2) {$t_{1}$};
\node at (0+3,-0.8) {$t_{2}$};
\node at (1.15/2+3,-0.2) {$t_{3}$};

\draw (0,0-2) circle (0.1) [fill=red];
\draw (-0.7,-1-2) circle (0.1) [fill=green];
\draw (0.7,-1-2) circle (0.1) [fill=blue];
\draw (-0.15,-0.15-2)--(-0.7,-0.85-2) [-latex];
\draw (0.55,-1-2)--(-0.55,-1-2) [-latex];
\draw (0.7,-0.85-2)--(0.15,-0.15-2) [-latex];
\node at (-1.15/2,-0.2-2) {$t_{1}$};
\node at (0,-0.8-2) {$t_{2}$};
\node at (1.15/2,-0.2-2) {$t_{3}$};

\draw (0+3,0-2) circle (0.1) [fill=red];
\draw (-0.7+3,-1-2) circle (0.1) [fill=green];
\draw (0.7+3,-1-2) circle (0.1) [fill=blue];
\draw (-0.15+3,-0.15-2)--(-0.7+3,-0.85-2) [-latex];
\draw (0.55+3,-1-2)--(-0.55+3,-1-2) [-latex];
\draw (0.15+3,-0.15-2)--(0.7+3,-0.85-2) [-latex];
\node at (-1.15/2+3,-0.2-2) {$t_{1}$};
\node at (0+3,-0.8-2) {$t_{2}$};
\node at (1.15/2+3,-0.2-2) {$t_{3}$};
\end{tikzpicture}
\caption{Some motif patterns in temporal networks, with $t_{1}<t_{2}<t_{3}$ and $t_{3}-t_{1}\leq \delta$. } 
\label{figure:motifexample}
\end{figure}

To desribe and compare temporal networks from an abstract level, motif feature was proposed as a promising option. 
In particular, we focus the motifs of influence propagation between two or three nodes and three edges as in \cite{paranjape2017motifs}, some examples are shown by Fig. \ref{figure:motifexample}. 
To form the motif feature of a temporal network, altogether 36 kinds of motifs within a temporal network are counted. 
With the first edge goes from $u$ to $v$, the second and the third edge can be any one from $(u\rightarrow v,v\rightarrow u, u\rightarrow w,w\rightarrow u, v\rightarrow w,w\rightarrow v)$, resulting in 36 possibilities. 
All edges in a motif have to be activated in a time window with width $\delta$. 
By dynamic programming, this counting procedure can be done efficiently. 
We adopted the implementation in \cite{paranjape2017motifs}. 

The motif feature is a promising representation for temporal networks since:
\begin{enumerate}
\item Motif feature has only 36 components for a temporal network of an arbitrary size. 
\item Real temporal networks are stable w.r.t. motif. 
For example, the motif feature of a traffic network from January to Febuary is very close to that from Febuary to March.
\item The motif features for temporal networks of different disciplines are distinguishable. 
Hence the difference in network structures or propagation models is captured by the motif feature. 
\end{enumerate}
Therefore, motif feature is arguably one candidate for temporal network fingerprint. 

\section{Motivation and Metrics}
\label{section:3}
\subsection{Motivation}
So far, the evaluation of diffusion models has been subjective and unprincipled.
Given the proposition that motif feature can capture the properties of real temporal networks, we conjecture that the generated temporal networks should have the similar characteristics as real ones regarding motif.
Denote the algorithm that generates the motif feature from a temporal network by $\texttt{m}$, a metric on motif feature space is denoted by $d$. 
The temporal network generated from network topology $G$, diffusion model $M$ with random seed $k$ after mixing independent sequences by Algo. \ref{algorithm:1} is denoted by $TG_{M,G,k}$. 
We are now ready to formalize the intuition.
Consider a diffusion model $M$ and some static networks $\mathcal{G}=\left\{G_{n}\right\}_{n=1}^{N}$. 
When $M$ is run on a specific $G\in\mathcal{G}$ for several times with random seeds $\mathcal{K}=\left\{k_{q}\right\}_{q=1}^{Q}$, the generated temporal networks 
$$\left\{TG_{M,G,k_{q}} \right\}_{q=1}^{Q}$$ 
should be similar w.r.t. the motif feature. 
When $M$ is run on different network skeletons, the generated temporal networks 
$$\left\{TG_{M,G_{n},k} \right\}_{n=1}^{N}$$ 
should be distinguishable w.r.t. motif. 

\subsection{Metrics}
Formally, a diffusion model $M$ should satistfy the following two conditions:

\begin{table*}[htbp]
\caption{Network structures adopted for experiments.}
\begin{center}
\begin{tabular}{c|ccccccc}
\toprule[1.5pt]
\textbf{Properties} & $G_{1}$\cite{richardson2003trust}.& $G_{2}$ \cite{leskovec2012learning}. & $G_{3}$ \cite{leskovec2010predicting}. & $G_{4}$ \cite{feather}. & $G_{5}$ \cite{rozemberczki2019multiscale}. & $G_{6}$ \cite{rozemberczki2019multiscale}. &  $G_{7}$ \cite{klimt2004introducing}. \\

\midrule[1pt]
No. of nodes. &75,879 & 4,039 & 7,115 & 7,624 & 22,470 & 37,700 & 36,692\\

\midrule
No. of edges. & 508,837 & 88,234 & 103,689 & 27,806 & 171,002 & 289,003 & 183,831 \\

\midrule
Discipline & \tabincell{c}{Social\\network.} & \tabincell{c}{Social\\network.} & \tabincell{c}{Social\\network.} & \tabincell{c}{Social\\network.} & \tabincell{c}{Social\\network.} & \tabincell{c}{Github\\network.} & \tabincell{c}{Email \\ network.} \\
\toprule

\textbf{Properties}& $G_{8}$ \cite{leskovec2007graph}. & $G_{9}$ \cite{leskovec2005graphs}. & $G_{10}$ \cite{leskovec2005graphs}. \\
\midrule
No. of nodes.  & 265,214 & 34,546 & 27,770 \\
\midrule
No. of edges. & 420,045 & 421,578 & 352,807 \\
\midrule
Discipline& \tabincell{c}{Email\\network.} & \tabincell{c}{Citation\\network.} & \tabincell{c}{Citation\\network.} \\
\bottomrule[1pt]
\end{tabular}
\label{table:1}
\end{center}
\end{table*}

\subsubsection{Stability}
For an arbitrary static network $G$, 
\begin{equation}
\label{equation:stability}
\begin{aligned}
\text{Pr}_{k_{1},k_{2}}&\left\{d(\texttt{m}(TG_{M,G,k_{1}}),\texttt{m}(TG_{M,G,k_{2}}))\leq \tau \right\}\\
& \geq 1-\epsilon,\\
\end{aligned}
\end{equation}
where $\tau$ reflects the tolerance to the fluctuation of the motif features.
Computing the probability in \eqref{equation:stability} requires integrating out the random seeds, which can be statistically approximated by exhausting the finite collection of $Q$ random seeds $\mathcal{K}$. 
The stability can also be measured from the magnitude of fluctuation w.r.t. the choice of random seeds, formally we define the \emph{stability score} of diffusion model $M$ w.r.t. static network $G$ (and implicitly random seeds set $\mathcal{K}$) as:
\begin{equation}
\label{equation:stability2}
\begin{aligned}
s^{\text{stab}}&(M,G)=\\
&\max_{k_{1},k_{2}\in \mathcal{K}}\left\{d(\texttt{m}(TG_{M,G,k_{1}}),\texttt{m}(TG_{M,G,k_{2}})) \right\},\\
\end{aligned}
\end{equation}
which, given $G$, is a monotonic function of $\mathcal{K}$.
If $s^{\text{stab}}(M,G)$ is smaller than $\tau$ for $\mathcal{K}$ complex enough then \eqref{equation:stability} is likely to hold. 
This definition can be made independent of network topology by exhausting all available network datasets in $\mathcal{G}$.
\begin{equation}
\label{equation:stability3}
s^{\text{stab}}(M)=\max_{G\in\mathcal{G}}\left\{s^{\text{stab}}(M,G) \right\}.
\end{equation}
The definition in \eqref{equation:stability3} is a measure of the variation of a diffusion model in the worst case, therefore it is a measure of stability. 
By boiling down \eqref{equation:stability} to \eqref{equation:stability3}, the stability of a diffusion model w.r.t. motif can be efficiently computed. 

\subsubsection{Separability}
For arbitrary $G_{1}\neq G_{2}$,
\begin{equation}
\label{equation:separability}
\text{Pr}_{k}\left\{d(\texttt{m}(TG_{M,G_{1},k}),\texttt{m}(TG_{M,G_{2},k}))\leq \tau \right\}\leq \epsilon,
\end{equation}
where the probability is computed by intergrating out the random seed. 
The definition in \eqref{equation:separability} is too strong since $G_{1}$ and $G_{2}$ might have a insignificance difference that does not influence the motif feature. 
Instead, $G_{1}$ and $G_{2}$ should be of different disciplines rather than variants of a third network. 
In practice, we collect network across a diversity of disciplines to form $\mathcal{G}$ to ensure the heterogeneity within. 
As in the reduction of stability, in practice, separability is more conveniently captured by: 
\begin{equation}
\label{equation:separability2}
\begin{aligned}
s^{\text{sep}}&(M)=\\
&\min_{G_{1},G_{2}\in\mathcal{G},k\in \mathcal{K}}\left\{d(\texttt{m}(TG_{M,G_{1},k}),\texttt{m}(TG_{M,G_{2},k})) \right\}.\\
\end{aligned}
\end{equation}

If $s^{\text{sep}}(M)$ is larger than $\tau$ then for $Q,N$ large enough then \eqref{equation:separability} is expected to hold. 

\subsection{Motif Distance}
To measure the similarity between diffusion models from their motif features, we define the \emph{motif distance} between $M_{1}$ and $M_{2}$ as:
\begin{equation}
\label{equation:distance}
d^{\text{m}}(M_{1},M_{2})=\frac{\sum_{n}^{N}d(TG_{M_{1},G_{n}},TG_{M_{2},G_{n}})}{N},
\end{equation}
where $TG_{M_{i},G_{n}}$ is the average motif feature over the choice of the random seed. 
Equation \eqref{equation:distance} can find similar diffusion models from the motif features they introduce. 
Moreover, \eqref{equation:distance} is an abstract and high-level metrics that cares not the detailed implementation of diffusion process.
But it would be very difficult to design a diffusion model that is motifly far from a given model, since motif feature yields not gradient. 

\section{Empirical observations}
\label{section:4}
\subsection{Settings}
To compute \eqref{equation:stability3} and \eqref{equation:separability2} for a given diffusion model $M$, we collected a set of $N=10$ static networks as $\mathcal{G}$, whose details are listed in Table \ref{table:1}.
Then a collection of $Q=50$ random seeds were sampled uniformly from range $[0,2^{20}-1]$ to form $\mathcal{K}$. 
The implementation in \cite{paranjape2017motifs} was adopted as the motif feature extractor $\texttt{m}$. 
All 36 components were normalized so they sum up to unity. 
As for $d$, we adopted:
\begin{equation}
d(m_{1},m_{2})=\sum_{i=1}^{36}(m_{1,i}-m_{2,i})^{2}.
\end{equation}
We evaluated and compared several common diffusion models including three mainstream diffusion models and three variants of the independent cascade model:
\begin{enumerate}
\item The independent cascade model $M_{\text{IC}}$\cite{saito2008prediction}.
\item The weighted cascade model $M_{\text{WC}}$\cite{hong2019seeds}.
\item The linear threshold model $M_{\text{LT}}$\cite{chen2010scalable}.
\item The message model $M_{\text{SM}}$, in which each participant only communicates to three-five friends in both directions.
\item The double confirmation model $M_{\text{DC}}$, in which the receiver of a message might send a confirmation back to the sender.
\item The bank model $M_{\text{BK}}$, in which after an edge is activated, the receiver of the message might communicate with a center node within the community. 
\end{enumerate}
Each cascade begins with ten activated nodes and the propagation continues for six rounds. 
A generated temporal network is the combination of $P=10$ cascades. 

\subsection{Stability and Separability}
The stability and separability of all six diffusion models were computed using \eqref{equation:stability3}\eqref{equation:separability2} with the settings above, the results are shown in Table \ref{table:2}.
\begin{table}[htbp]
\caption{Stability and separability of diffusion models.}
\begin{center}
\begin{tabular}{c|cc}
\toprule[1.5pt]
\tabincell{c}{\textbf{Diffusion}\\\textbf{model.}} & \tabincell{c}{Stability. \\ $s^{\text{stab}}$} & \tabincell{c}{Separability. \\ $s^{\text{sep}}$} \\

\midrule[1pt]
$M_{\text{IC}}$ & 0.001465 & 0.041895\\
$M_{\text{WC}}$ & 0.001478 & 0.004188\\
$M_{\text{LT}}$ & 0.031853 & 0.043319\\
$M_{\text{SM}}$ & 0.000042 & 0.001036\\
$M_{\text{DC}}$ & 0.000713 & 0.003169\\
$M_{\text{BK}}$ & 0.000405 & 0.001720\\
\bottomrule[1pt]
\end{tabular}
\label{table:2}
\end{center}
\end{table}

For all diffusion models there was $s^{\text{sep}}<s^{\text{stab}}$. 
This indicates that for any model, the fluctuation between the motif features of different network topologies is larger than that in one network caused by randomness inside the model. 

Given a diffusion model $M$ with 
\begin{equation}
s^{\text{sep}}(M)<s^{\text{stab}}(M), 
\end{equation}
and a temporal network instance $TG_{G,M,k}$, one can distinguish $TG_{G,M,k'}$ from $TG_{G',M,k}$ by computing the distances. 
All compared models satistify this condition. 

To interpret from a machine learning perspective, $s^{\text{stab}}$ can be understood as the maximum noise within a class and $s^{\text{sep}}$ is the minimal distance between the centroids of two classes. 
So $s^{\text{sep}}(M)<s^{\text{stab}}(M)$ means that the classification task is very easy and confusions hardly appear. 

\subsection{Motif Distance}
\begin{figure}[htbp]
\centering
\includegraphics[width=0.4\textwidth] {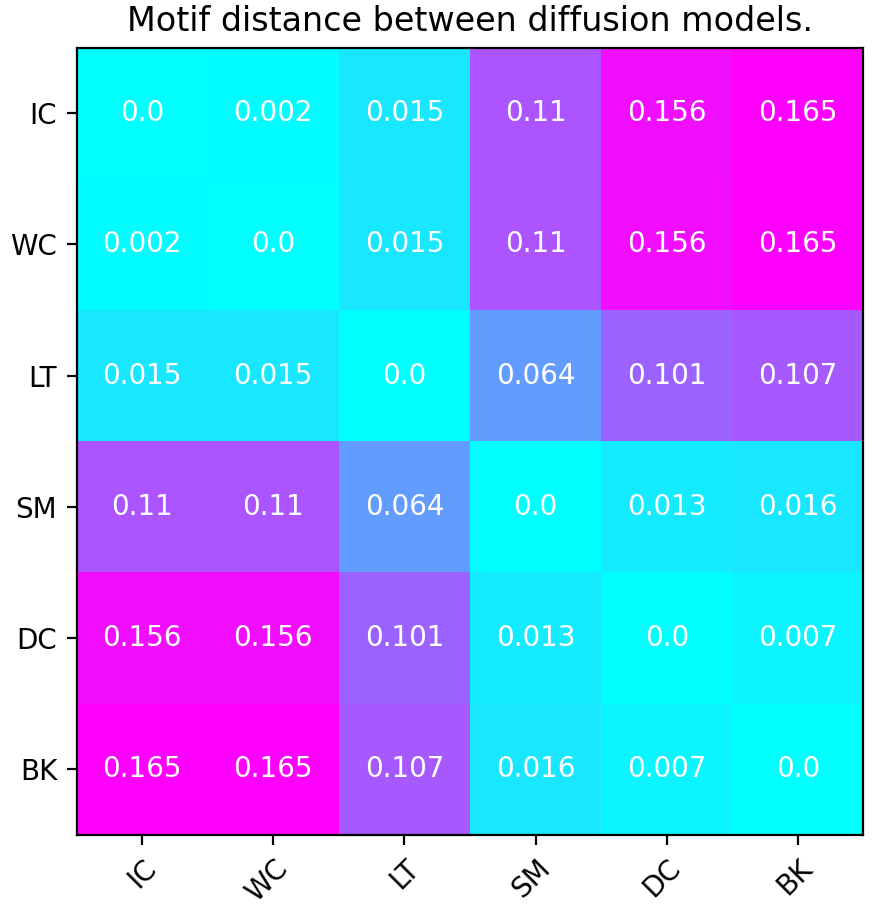}
\caption{The motif distance $d^{\text{m}}$ between diffusion models.} 
\label{figure:motifdistance}
\end{figure}

The distances between all six diffusion models were collected in Fig. \ref{figure:motifdistance}

One problem is that the distance between some pairs of diffusion models turns out to be too small. 
For example, the distance between $M_{\text{IC}}$ and $M_{\text{WC}}$ is smaller than $s^{\text{sep}}(M_{\text{IC}})$. 
That is to say, it is almost always that the distance between $TG_{M_{\text{IC}},G,k}$ and $TG_{M_{\text{WC}},G,k}$ is smaller than that between $TG_{M_{\text{IC}},G,k}$ and $TG_{M_{\text{IC}},G',k}$. 

If a pair of diffusion models has a small motif distance then it is arguably that they yield similar propagation pattern and are interchangable in practice. 

\subsection{Discussion}
\begin{figure}[htbp]
\centering
\subfigure[$M_{\text{IC}}$ on $G_{1}$.]{
\begin{minipage}[c]{0.23\textwidth}
\includegraphics[width=3.8cm]{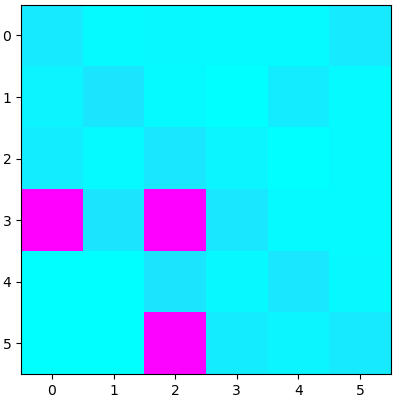}
%\caption{fig1}
\end{minipage}%
}%
\subfigure[$M_{\text{IC}}$ on $G_{6}$.]{
\begin{minipage}[c]{0.23\textwidth}
\includegraphics[width=3.8cm]{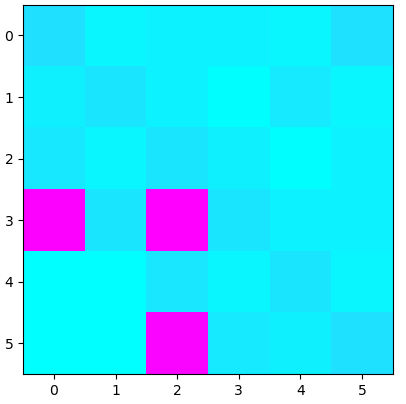}
%\caption{fig2}
\end{minipage}%
}%     

\subfigure[$M_{\text{IC}}$ on $G_{7}$.]{
\begin{minipage}[c]{0.23\textwidth}
\includegraphics[width=3.8cm]{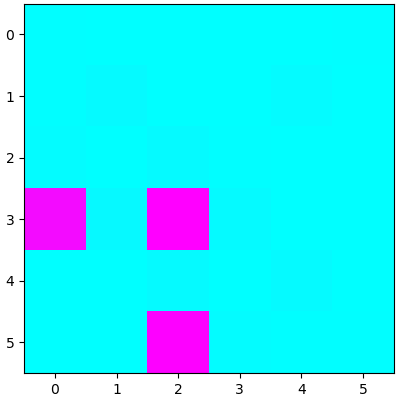}
%\caption{fig1}
\end{minipage}%
}%
\subfigure[$M_{\text{IC}}$ on $G_{10}$.]{
\begin{minipage}[c]{0.23\textwidth}
\includegraphics[width=3.8cm]{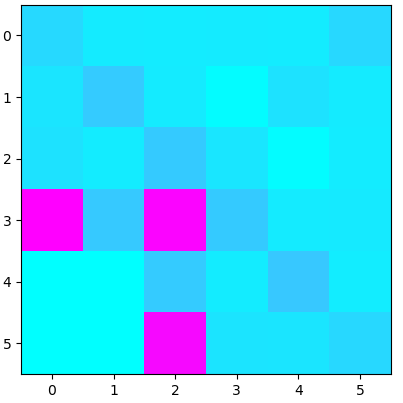}
%\caption{fig2}
\end{minipage}%
}%  

\subfigure[$M_{\text{DC}}$ on $G_{1}$.]{
\begin{minipage}[c]{0.23\textwidth}
\includegraphics[width=3.8cm]{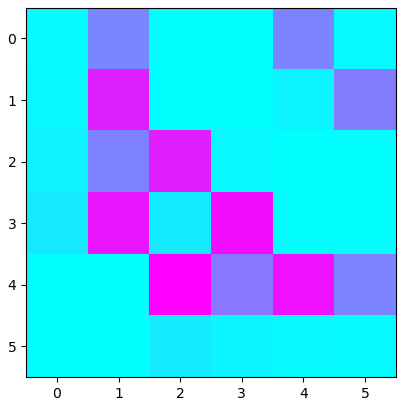}
%\caption{fig2}
\end{minipage}%
}%
\subfigure[$M_{\text{DC}}$ on $G_{6}$.]{
\begin{minipage}[c]{0.23\textwidth}
\includegraphics[width=3.8cm]{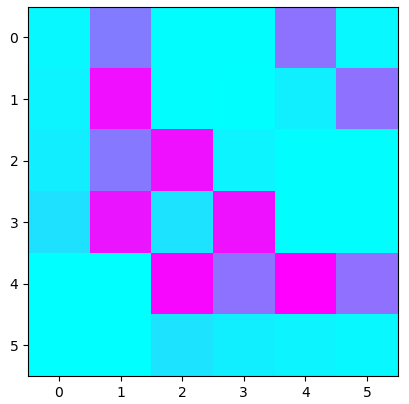}
%\caption{fig1}
\end{minipage}%
}%

\subfigure[$M_{\text{DC}}$ on $G_{7}$.]{
\begin{minipage}[c]{0.23\textwidth}
\includegraphics[width=3.8cm]{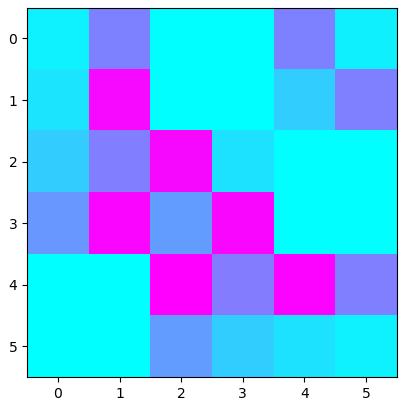}
%\caption{fig2}
\end{minipage}%
}%
\subfigure[$M_{\text{DC}}$ on $G_{10}$.]{
\begin{minipage}[c]{0.23\textwidth}
\includegraphics[width=3.8cm]{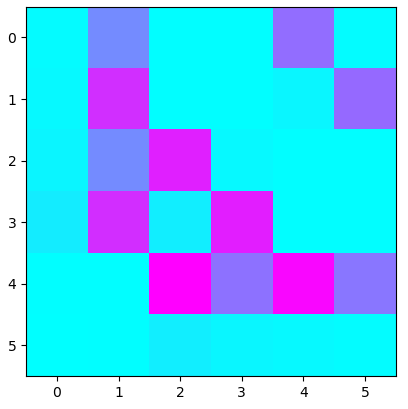}
%\caption{fig1}
\end{minipage}%
}%
\centering
\caption{Examples of motif features.}
\label{figure:motifs}
\end{figure}

Although we have the empirical observation that for an arbitrary listed diffusion model $M$ the condition $s^{\text{sep}}(M)<s^{\text{stab}}(M)$ holds, the inter-network distance $s^{\text{stab}}(M)$ is still very small. 
As an example, the motif feature of $M_{IC}$ and $M_{DC}$ on $G_{1},G_{6},G_{7},G_{10}$ are collected in Fig. \ref{figure:motifs}

The motif features of $M_{\text{WC}}$ and $M_{\text{LT}}$ are similar to those of $M_{\text{IC}}$, while those of $M_{\text{SM}}$ and $M_{\text{BK}}$ are similar to $M_{\text{DC}}$, a fact that has been demonstrated by Fig. \ref{figure:motifdistance}. 
 
Fig. \ref{figure:motifs}(a)-(d) and Fig. \ref{figure:motifs} (e)-(h) reflect two representative motif feature patterns in all six diffusion models (six diffusion models, but only two representative patterns). 
The first pattern in Fig. \ref{figure:motifs}(a)-(d) is also that of the real temporal network Email-EU\cite{paranjape2017motifs}\cite{klimt2004introducing}. 
While the second pattern in Fig. \ref{figure:motifs}(e)-(h) is more similar to that of a real short message temporal network \cite{paranjape2017motifs}.

It can now be concluded that it is the diffusion model rather than the static network structure that dominates the motif features. 
The implication of this assertation is that the diversity of motif features of temporal networks mainly origins from the diversity in propagation pattern rather than the network topology. 
This fact also calls for designing more diffusion model for specific scenario/disciplines. 
The mostly adopted diffusion models, $M_{\text{IC}}$, $M_{\text{WC}}$, $M_{\text{LT}}$ yield only one diffusion model since the motif distances between any pair of them is small. 
But this pattern is insufficient for covering the diversified propagation patterns observed in real temporal networks. 
Such insufficiency is not relieved even GNN is introduced into network analysis, since GNN only simulates the propagation pattern of the classical diffusion models.  
Only after new diffusion models are proposed can this insufficiency be finally solved. 

Moreover, our empirical observation challenges those studies that built their generality on the diversity of network datasets instead of that of diffusion models. 

\section{Conclusion}
\label{section:5}
In this paper we borrow the idea of the motif feature to quantitively evaluate and compare diffusion models. 
The three components in generating an artificial temporal network: the diffusion model, the network topology and the randomness are decomposed in order to measure their weights in forming a motif feature.  
Two metrics, stability and separability are defined from motif to evaluate a diffusion model. 
We reduce these two definitions to computable forms and evaluate them for various diffusion models. 

Moreover, we proposed a motif-based distance metrics between diffusion models. 
Comprehensive results show that although most diffusion models are stable and successfully separate different networks, the distance between them appears to be very small.
That is to say, the current mainstream diffusion models fail to provide enough diversity in modelling temporal influence propagation. 
As a result, more scenario-specific diffusion models have to be designed for more realistic simulations in general network analysis.

\section*{Acknowledgment}
The author received advices from Dr. Wenwen Xia, Dr. Chong Di, and Runbo Ni. 

Abaaba.

\bibliographystyle{ieeetr}
\bibliography{diff.bib}

\end{spacing}
\end{document}